\documentclass[12pt,draftcls,onecolumn]{IEEEtran}

\usepackage[english]{babel}

\usepackage[cmex10]{amsmath}
\usepackage{epsfig}
\usepackage[utf8]{inputenc}
\usepackage{amssymb}
\usepackage{color}

\usepackage{dblfloatfix}
\usepackage{cite}

\addto\captionsenglish{%
  \renewcommand{\figurename}{Fig.}%
}

\newcounter{MYtempeqncnt}

\newcommand{ \pr }{ \mathbb{P} }
\newcommand{ \rP }{ R }
\newcommand{ \rPown }{ {\rP_0} }
\newcommand{ \rPii }{ {\rP_{i}} }
\newcommand{ \nP }{ \sigma_\text{n}^2 }
\newcommand{ \ff }{ \mathbf{H} }
\newcommand{ \ffOwn }{ \ff_{0} }
\newcommand{ \ffIi }{ \ff_{i} }
\newcommand{ \eqV }{ \mathbf{h}_{\text{eq}} }
\newcommand{ \eqVi }{ \eqV^{(i)} }
\newcommand{ \fEl }{ h }
\newcommand{ \fElOwn }{ \fEl }
\newcommand{ \fElInt }{ g }
\newcommand{ \nR }{ {N_\text{R}} }
\newcommand{ \nT }{ {N_\text{T}} }
\newcommand{ \nL }{ {N_\text{L}} }
\newcommand{ \nLi }{ \nL^{(i)} }
\newcommand{ \pE }{ w }
\newcommand{ \pV }{ \mathbf{\pE} }
\newcommand{ \pVown }{ \pV_0 }
\newcommand{ \pM }{ \mathbf{W} }
\newcommand{ \rEl }{ r }
\newcommand{ \rElOwn }{ \overline{\rEl} }
\newcommand{ \rElInt }{ \tilde{\rEl} }
\newcommand{ \rV }{ \mathbf{\rEl} }
\newcommand{ \rVown }{ \overline{\rV} }
\newcommand{ \rVint }{ \tilde{\rV} }
\newcommand{ \sEst }{ \hat{\rV} }
\newcommand{ \dE }{ d }
\newcommand{ \dEown }{ \dE_0 }
\newcommand{ \dV }{ \mathbf{\dE} }
\newcommand{ \nV }{ \mathbf{n} }

\newcommand{ \mEig }{ {\lambda_\text{max}} }
\newcommand{ \rat }{ \psi }
\newcommand{ \ratOwn }{ \rat_{0} }

\newcommand{ \numV }{ x }
\newcommand{ \denV }{ y }
\newcommand{ \eigC }{ \varphi }
\newcommand{ \pOut }{ {p_\text{out}} }
\newcommand{ \sinr }{ \gamma }
\newcommand{ \rF }{ \mathbf{F} }
\newcommand{ \cc }{ \star }
\newcommand{ \herm }{ \dagger }
\newcommand{ \minL }{ M }
\newcommand{ \maxL }{ N }
\newcommand{ \nInt }{ K }
\newcommand{ \nGrp }{ {p'} }
\newcommand{ \nMem }{ t' }
\newcommand{ \expC }{ b }
\newcommand{ \tuple }{ \theta }
\newcommand{ \sOne }{ \Lambda }
\newcommand{ \stbcSone }{ \Omega }
\newcommand{ \stbcStwo }{ \Phi }
\newcommand{ \stbcSthree }{ \Upsilon }

\begin{document}

\title{Higher Rank Interference Effect on Weak Beamforming or OSTBC Terminals}

\author{
Michal~\v{C}ierny,
Zhi~Ding,~\IEEEmembership{Fellow,~IEEE,}
Risto~Wichman%
\thanks{This work has been submitted to the IEEE for possible publication.  Copyright may
be transferred without notice, after which this version may no longer be accessible.}%
\thanks{This work was supported by Academy of Finland, National Science Foundation
grants 1147930 and 1321143, TEKES, Graduate School in Electronics, Telecommunications
and Automation (GETA), HPY Foundation and Nokia Foundation.}
\thanks{M. \v{C}ierny is with Nokia Networks, 02610 Espoo, Finland
(e-mail: michal.cierny@nsn.com).}
\thanks{Z. Ding is with Department of Electrical and Computer Engineering, University
of California, Davis, California 95616 (e-mail: zding@ucdavis.edu).}
\thanks{R. Wichman is with Department of Signal Processing and
Acoustics, Aalto University School of Electrical Engineering, 00076 Aalto, Finland
(e-mail: risto.wichman@aalto.fi).}
}

\maketitle

\begin{abstract}
User performance on a wireless network depends on whether a neighboring cochannel
interferer applies a single (spatial) stream or a multi stream transmission.
This work analyzes the impact of interference rank on a beamforming and 
orthogonal space-time block coded (OSTBC) user transmission. 
We generalize existing analytical results on signal-to-interference-plus-noise-ratio (SINR)
distribution and outage probability under arbitrary number of unequal power interferers. 
We show that higher rank interference causes lower outage probability, 
and can support better outage threshold especially in the case of beamforming.
\end{abstract}


\section{Introduction}
\label{sec:intro}

\IEEEPARstart{M}{ulti-antenna} transmission techniques \cite{BiGoPaBook} have 
substantially transformed the modern wireless communications 
by providing effective diversity means for improving wireless network capacity 
and radio link reliability. Modern cellular systems, such
as 3GPP Long Term Evolution (LTE) \cite{DaSaSkBeBook}, extensively rely on multiple-input and
multiple-output (MIMO) techniques.

Interference is inherently a major capacity limiting factor in cellular networks
and comes in quite a few shapes: intra-cell interference, inter-cell interference, 
interference between spatial streams for the same user, etc. 
In this work we focus on inter-cell interference, i.e., 
interference between adjacent cells, also known as other cell interference (OCI). 
Though it is well known in the community that spatial multiplexing is susceptible to OCI
\cite{AnChHe2007}, the converse has not been explored well. 
In other words, there is little information on what effect spatially multiplexed interference 
may have on other types of transmissions within the system. 

Our study is motived by situations when a receiver with a weak desirable signal is interfered
by another transmitter that may have the option to choose the rank of its
transmission for sending single-layer or multi-layer MIMO
transmissions. Having such choices requires the interferer to have a strong channel to its
own receiver as spatial multiplexing usually does not fare well in low
signal-to-interference-plus-noise-ratio (SINR) regime.
This can happen when a user equipment (UE) is near the cell edge and is being
interfered by a neighbor base station (BS) serving other UEs on a strong link. In a
more detailed example, a
macrocell UE may be located within a coverage blind spot around a closed access femtocell
\cite{ChAnGa2008}. In a femtocell, the short link between a transmitter and a receiver tends
to be strong and chances to use spatial multiplexing can be high.
  
The way we approached our analysis is similar to \cite{Ah2009} which
studied the performance of beamforming. The work in \cite{Ah2009} does not consider
interferers performing spatial multiplexing or orthogonal space-time block coding
(OSTBC) transmissions. We found one contribution
to such mixed MIMO cases in \cite{RaCaPr2007} where the authors simulated a hexagonal
cellular network layout and collected SINR as well as bit error rate statistics. 
Another related work appears in \cite{LiZhCiZh2011}, which only considers
signal-to-interference ratio (SIR) without channel noise and does not tackle outage
probability. Performance of OSTBC under various MIMO interference has been analyzed
in \cite{LiCiHi2008,LiZhCiZh2011_2,ZhLiCi2012}. Similar to \cite{LiZhCiZh2011} the authors
analyzed SIR distributions with neither noise nor outage probability. A simulation study that
included antenna correlation is given in \cite{JiWuLi2012}.
  
The difference between spatial multiplexing and beamforming transmissions
is reflected in the rank of interference signal space. 
In this paper we analyze the impact of interference rank on the performance
of a receiver whose own transmitter also applies beamforming or OSTBC. 
We consider 
arbitrary number of interferers, each of whom with their own transmission power and
multi-antenna technique. 
We place no limit on the number of antennas at the 
transmitter or the receiver.
We incorporate realistic channel conditions
including near-static long term component and short term Rayleigh fading.
We derive a closed-form outage probability, verify its validity using Monte
Carlo simulations, and use it to assess the impact of interference rank on the
own transmission. 
  
Our results suggest that higher rank transmission of a strong interferer has a lower
probability of causing outage than a rank~1 interferer.  When the desired user transmitter
performs beamforming, this translates to more than $2\mathrm{dB}$ gain in the
supported SINR threshold for high dimension MIMO receivers. 
With the user signal transmitter performing OSTBC, the gain is still apparent but 
drops below $1\mathrm{dB}$.

We specify three main contributions in our work:
\begin{itemize}
  \item We directly show the effect of interference rank on a single
    layer MIMO transmission. We derive  SINR and outage
    probability for different interference ranks. Although our analysis
    is built on existing results \cite{Ah2009}, this is the first known comprehensive study on
    interference rank.
  \item We extend the known results on how interferers with arbitrary
    multi-antenna transmission techniques affect beamforming and OSTBC
    transmission \cite{ZhLiCi2012} by including noise power,
    as well as by deriving the
    probability of outage as in \cite{Ah2009}. Hence, our analysis is not
    limited to interference limited scenarios and, for given outage threshold, 
    provides a clearer performance metric.
  \item We offer a better insight into how precoded interference affects OSTBC own
    transmission than previously known \cite{LiCiHi2008,LiZhCiZh2011_2,ZhLiCi2012}
    by deriving the mean value of the random interference, thereby
    justifying existing approximation.
\end{itemize}
Our paper is organized as follows. Following this introduction,
Section \ref{sec:sysMod} summarizes the system model with major assumptions. Then,
in Section \ref{sec:anl} we derive the SINR distribution and probability of outage
of beamforming and OSTBC under arbitrary number of interferers that perform OSTBC or
precoding. In Section \ref{sec:resDis} we validate our SINR and outage probability
to present our main results and discuss their impact. Finally Section
\ref{sec:outro} concludes our manuscript.

\section{System Model}
\label{sec:sysMod}

\begin{figure}[t]
  \begin{center}
    \includegraphics[scale=0.23]{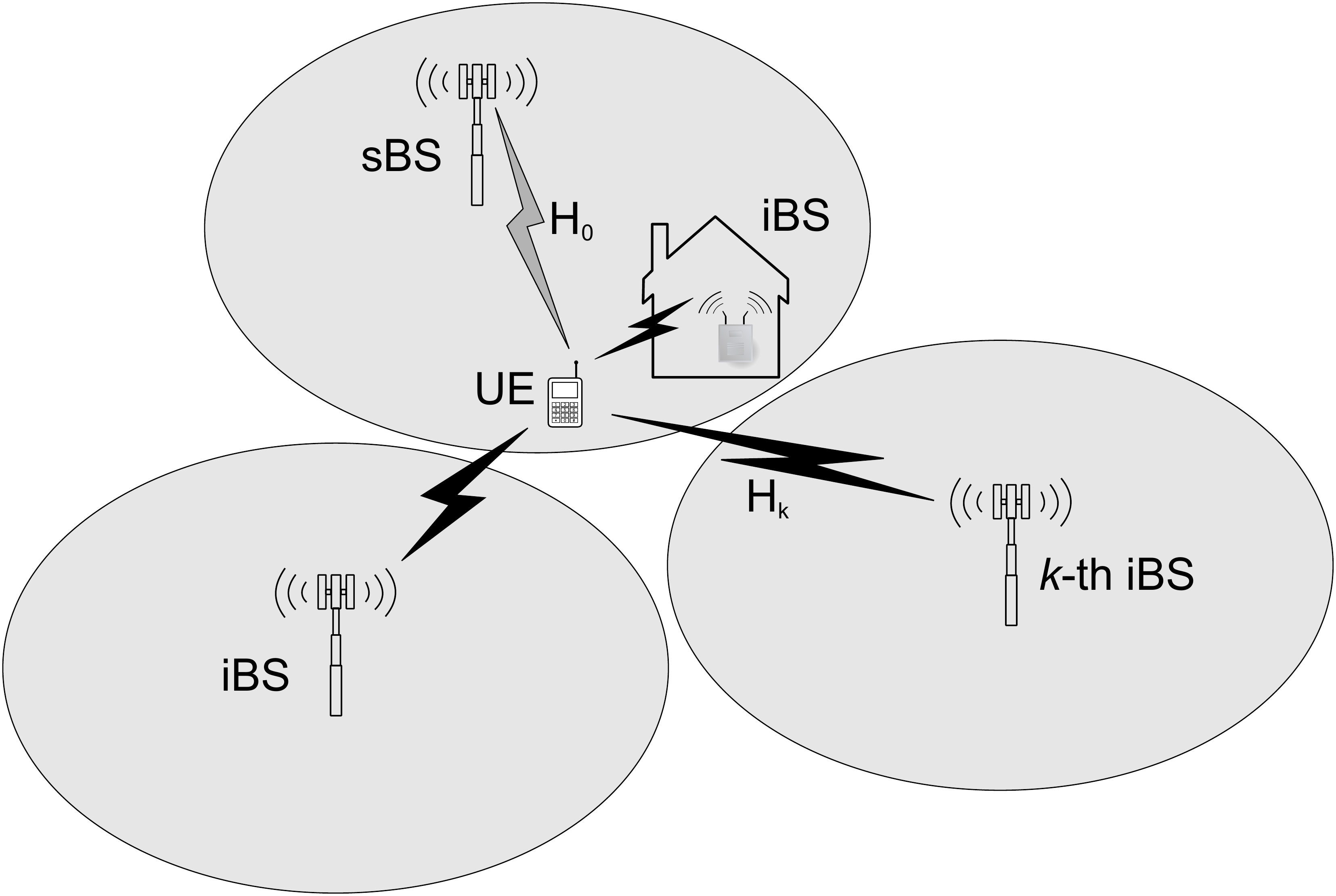}
    \caption{An example scenario with an UE receiving its own signal from a macro sBS
      and interference signals from a femto iBS and two macro
      iBSs.}\label{fig:scen}
  \end{center}
\end{figure}

Consider a general cellular downlink scenario with a UE that receives its own signal
from a single serving base station (sBS) and interfering signals from arbitrary number
$\nInt$ of interfering base stations (iBSs). An
illustrative scenario with a femto iBS and several macro iBSs is shown in
\figurename~\ref{fig:scen}.
The radio
channel between any link from a BS to an UE consists of a long term component, typically
depending on large-scale channel models on pathloss and shadowing effect, and a small-scale
Rayleigh fading component. We analyze quasi-static situations 
where the large scale channel parameters are constant across time and spatial
subchannels, where considering the average effect of fast fading components
that are flat in frequency but vary for each time instant and spatial subchannel.
The spatial components of fast fading are assumed to be i.i.d.

Consider that the UE has $\nR$ receive antennas and each of the BSs has $\nT$ transmit
antennas. We
denote the long term received power of the user signal as $\rPown$ and the long term
received power of $i$-th iBS as $\rPii$. Values $\rPown$ and $\rPii$ contain all
coupling gain components (transmission power, long term channel effect, noise figure,
etc.) except the fast fading channel effect. The fast fading component between the UE
and sBS is a $\nR \!\times\! \nT$ matrix $\ffOwn$ with complex Gaussian elements with
zero mean and unit variance. The fast fading component between
the UE and $i$-th iBS is a same type $\nR \!\times\! \nT$ matrix and is denoted by
$\ffIi$.
Received signal at the UE is further corrupted by additive white Gaussian noise with
power $\nP$. For simplicity, we normalize the data symbols to be of unit energy.

We consider two transmission schemes in this work. In situations when a BS has reliable
channel state information (CSI), it shall apply precoding, i.e., a closed-loop MIMO
transmission. As focus is to assist a
UEs with weak user signals (lower SINR regime), the sBS will be restricted to single-layer
transmission, i.e., beamforming. The iBSs are not restricted in this way.
The number of transmission layers of $i$-th iBS, denoted as $\nLi$, can be
$1\le \nLi \le \min \{ \nR , \nT \}$.
For tractability purposes we assume that sBS
performs optimal beamforming based on eigen-decomposition of $\ffOwn^\herm \ffOwn$. This
is an ideal version of the codebook approach present in LTE. The receiver shall have
accurate CSI from channel estimation and perform maximum ratio combining (MRC) reception.

In situations when BS does not have reliable CSI with respect to
its downlink channel, it can utilize  OSTBC, 
where $\nT$ data symbols are encoded over $\nT$ time
instances. While it is known \cite{TaJaCa99} that a full rate OSTBC exists only for
$\nT \!=\! 2$, we can extrapolate the results to higher $\nT$ for illustrative 
purposes. If sBS performs OSTBC, UE applies coherent OSTBC receiver processing as
shown in \cite{Al98} and formalized in \cite{TaJaCa99}. In case an iBS performs OSTBC
its $\nLi$ is, naturally, equal to one.

\section{Analysis}
\label{sec:anl}

In this section we will analyze performance 
of beamforming and OSTBC under a finite
number of interferers that perform precoding or OSBTC. 
Our ultimate performance measure
shall be outage probability $\pOut$ defined as
\begin{equation}
  \pOut \triangleq \pr \left\{ \sinr \leq \sinr_0 \right\} = \int_0^{\sinr_0} p_\sinr
    (x) dx,
\end{equation}
where $\sinr$ is post-processing SINR and $\sinr_0$ is the outage threshold. 
Throughout the paper we will refer to post-processing SINR simply as SINR.

For both sBS transmission modes the SINR is expressed as
\begin{equation}
  \label{eq:sinrFrac}
  \sinr = \frac{\numV}{\denV +1},
\end{equation}
where $\numV$ represents received user signal power normalized by
noise power and $\denV$ represents received interference power normalized by
noise power. Our main task is then to find distributions of independent
random variables (RVs) $\numV$ and $\denV$. Once
we find them, we can calculate the probability density function (PDF) of SINR from
\begin{equation}
  \label{eq:sinrPdf}
  p_\sinr ( \sinr ) = \int_0^\infty (\denV+1) p_\numV ((\denV+1)\sinr ) p_\denV (\denV)
    d\denV
\end{equation}
and outage probability from
\begin{equation}
  \label{eq:pOut}
  \pOut = \int_0^{\sinr_0} \int_0^\infty (\denV+1) p_\numV ((\denV+1)\sinr ) p_\denV
    (\denV) d\denV d\sinr .
\end{equation}
Because our analysis is built in the same way as in \cite{Ah2009} and the references
within, one can also use distributions of $\numV$ and $\denV$ to approximate symbol error
rate of some specific modulation formats. However, as this step could not be considered
novel, we will leave it out of this work.

\subsection{Outage Probability of Beamforming}
\label{ssec:BF}

A received sample vector $\rV$ at the UE antenna ports can be expressed as
\begin{equation}
  \rV = \sqrt{\rPown} \ffOwn \pVown \dEown
  + \sum_{i=1}^\nInt \sqrt{\rPii} \eqVi + \nV,
\end{equation}
where $\pVown$ is a $\nT \!\times\! 1$ sBS precoding vector with unit Frobenius norm,
$\dEown$ is
sBS data symbol, $\eqVi$ is $\nR \!\times\! 1$ equivalent channel vector of the $i$-th
interferer and $\nV$ is $\nR \!\times\! 1$ noise sample vector. The insides of $\eqVi$
depend on transmission technique of the $i$-th iBS. Assuming $j$-th iBS performing
beamforming and $k$-th iBS performing spatial multiplexing, we have
\begin{equation}
  \eqV^{(j)} = \ff_j \pV_j \dE_j ,
\end{equation}
\begin{equation}
  \eqV^{(k)} = \ff_k \pM_k \dV_k = \sum_{m=1}^{\nL^{(k)}} \ff_k \pV_{km} \dE_{km}
    = \sum_{m=1}^{\nL^{(k)}} \eqV^{(km)} ,
\end{equation}
where $\pM_k$ is a $\nT \!\times\! \nL^{(k)}$ precoding matrix with unit Frobenius
norm and $\dV_k$ is $\nL^{(k)} \!\times\! 1$ symbol vector. With $k$-th iBS performing
spatial
multiplexing we further divide the equivalent channel $\eqV^{(k)}$ into contributions
from separate transmission layers $\eqV^{(km)}$, $m \!\in\! [ 1 , \nL^{(k)} ]$. In a
case when $l$-th iBS performs OSTBC the equivalent channel vector could be for example
\begin{equation}
  \label{eq:eqVecStbc}
  \eqV^{(l)} = \frac{1}{\sqrt{\nT}} \ff_l \dV_l ,
\end{equation}
where $\dV_l$ is $\nT \!\times\! 1$ symbol vector. Depending on the time instance, the
insides of $\eqV^{(l)}$ could also be a little different. However, that does not alter
the derivations that will follow. We also note here that the presented options may
cover other iBS transmission techniques, for example open loop spatial multiplexing
\cite{ZhLiCi2012}.

Using MRC filter $\pVown^\herm \ffOwn^\herm$ we define the SINR as
\begin{equation}
  \label{eq:bfSinrDef}
  \sinr \triangleq \frac { \rPown \big\Vert \pVown^\herm \ffOwn^\herm \ffOwn \pVown
    \big\Vert^2 }
    { \sum_{i=1}^\nInt \sum_{j=1}^{\nL^{(i)}} \rPii \big\Vert \pVown^\herm
    \ffOwn^\herm
    \eqV^{(ij)} \big\Vert^2 + \big\Vert \pVown^\herm \ffOwn^\herm \big\Vert^2 \nP } .
\end{equation}
As sBS uses optimal beamforming we get $\big\Vert \pVown^\herm \ffOwn^\herm
\big\Vert^2 \!=\! \mEig$, where $\mEig$ is the dominant eigenvalue of
$\ffOwn^\herm \ffOwn$. Dividing the SINR expression from \eqref{eq:bfSinrDef} by
$\mEig \nP$ we get the necessary shape as in \eqref{eq:sinrFrac}. The numerator RV
$\numV$ is given by
\begin{equation}
  \numV = \ratOwn \mEig,
\end{equation}
where $\ratOwn$ represents the long term SNR $\rPown / \nP$. Distribution of $\numV$
has been found in \cite{DiMaJa2003} and can be expressed as
\begin{equation}
  \label{eq:pdfBfX}
  p_\numV (\numV) = \sum_{k=1}^\minL
    \sum_{l=\maxL-\minL}^{(\maxL+\minL-2k)k} \!\!\!\!\!\! \eigC_{kl} \frac{\numV^l}
    {\Gamma (l+1)} \! \left( \frac{k}{\ratOwn} \right)^{l+1} \!\! e^{-\frac{\numV k}
    {\ratOwn}},
\end{equation}
where $\minL \!=\! \min \{ \nR,\nT \}$, $\maxL \!=\! \max \{ \nR,\nT \}$ and
$\eigC_{kl}$ are weight coefficients defined by
\begin{equation}
  \eigC_{kl} = \frac{ l! c_{kl} }
  {k^{l+1} \prod_{s=1}^\minL (\minL-s)! (\maxL-s)! },
\end{equation}
where $c_{kl}$ ensures that $\sum_{k=1}^\minL \sum_{l=\maxL-\minL}^
{(\maxL+\minL-2k)k} \eigC_{kl} \!=\! 1$. Values of $\eigC_{kl}$ can be found by
symbolic or numeric software. For the most common antenna configurations they
have been tabulated in \cite{DiMaJa2003}.

\begin{figure*}[!t]
  \normalsize
  \setcounter{MYtempeqncnt}{\value{equation}}
  \setcounter{equation}{18}
  \begin{equation}
    \label{eq:sinrPdfBf}
    f_\sinr (\sinr) =
    \sum_{i=1}^\nGrp
    \sum_{j=1}^{\nMem_i}
    \sum_{k=1}^\minL
    \sum_{l=\maxL-\minL}^{(\maxL+\minL-2k)k}
    \expC_{ij}
    \eigC_{kl}
    \sinr ^l
    e^{-\frac{k \sinr}{\ratOwn}}
    \sum_{r=0}^{l+1}
    \binom{l+1}{r}
    \frac{\Gamma (r+\nMem_i)}{l! \Gamma (\nMem_i)}
    \left( \frac{k}{\ratOwn} \right)^{l+1}
    \left( \frac{1}{\rat_i} \right)^j
    \left(
      \frac{\ratOwn}{k \sinr + \sOne_i}
    \right)^{r+j}
\end{equation}
  \begin{equation}
    \label{eq:pOutBf1}
    \pOut = \sum_{k=1}^\minL \sum_{l=\maxL-\minL}^{(\maxL+\minL-2k)k} \eigC_{kl}
      \left( 1 - e^{-\frac{k \sinr_0}{\ratOwn}}
      \left( \frac{1}{\rat_1} \right)^{\nMem_1}
	  \sum_{r=0}^l \sum_{s=0}^r
      \binom{r}{s} \frac{\Gamma (s + \nMem_1)}{r! \Gamma (\nMem_1)} \left(
      \frac{k \sinr_0}{\ratOwn} \right)^r \left( \frac{k \sinr_0}{\ratOwn} +
      \frac{1}{\rat_1} \right)^{-(s+\nMem_1)} \right)
  \end{equation}
  \begin{equation}
    \label{eq:pOutBf}
    \pOut = \sum_{i=1}^\nGrp \sum_{j=1}^{\nMem_i}
      \sum_{k=1}^\minL \sum_{l=\maxL-\minL}^{(\maxL+\minL-2k)k}
      \expC_{ij} \eigC_{kl} \left( 1 - e^{-\frac{k \sinr_0}{\ratOwn}}
      \left( \frac{\sOne_i}{k \sinr_0 + \sOne_i} \right)^j
      \sum_{r=0}^l \sum_{s=0}^r \binom{r}{s}
      \frac{\Gamma (s+j)}{r! \Gamma (j)}
      \left( \frac{k \sinr_0}{\ratOwn} \right)^r
      \left( \frac{\ratOwn}{k \sinr_0 + \sOne_i} \right)^s \right)
  \end{equation}
  \setcounter{equation}{\value{MYtempeqncnt}}
  \hrulefill
  \vspace*{0pt}
\end{figure*}

Let us now look at the structure of $\denV$. Firstly, with $k$-th iBS performing
beamforming or spatial multiplexing, vector $\ff_k \pV_{km} \dE_{km}$
has complex Gaussian elements with zero
mean and variance $1 / \nL^{(k)}$. This stems from the fact that matrices $\pM_k$
and vectors $\pV_k$ are normalized \cite{ShHa2000}. Similarly, multiplying given
term with $\pVown^\herm \ffOwn^\herm / \big\Vert \pVown^\herm \ffOwn^\herm
\big\Vert^2$ does not change Gaussianity of the elements and thus the whole term
$\big\Vert \pVown^\herm \ffOwn^\herm \eqV^{(km)} \big\Vert^2 / \big\Vert
\pVown^\herm \ffOwn^\herm \big\Vert^2$ is exponentially distributed with rate
$\nL^{(k)}$. Secondly, with $l$-th iBS performing OSTBC, $\ff_l$ in
\eqref{eq:eqVecStbc} is multiplied by normalized $\dV_l / \sqrt{\nT}$ leading to
the term $\big\Vert \pVown^\herm \ffOwn^\herm \eqV^{(l)} \big\Vert^2 / \big\Vert
\pVown^\herm \ffOwn^\herm \big\Vert^2$ being exponentially distributed with rate
$\nT$.


Variable $\denV$ is consequently given by a sum of weighted exponential RVs. The
weights $\rat_k$ and $\rat_l$, corresponding to precoding and OSTBC, respectively,
are given by
\begin{align}
  \rat_k &= \frac{\rP_k}{\nL^{(k)} \nP} , \\
  \rat_l &= \frac{\rP_l}{\nT \nP} .
\end{align}
The number of summed exponential RVs is $\sum_{m=1}^\nInt \nL^{(m)}$. We
can divide the contributions into $\nGrp$ groups with $i$-th group having $\nMem_i$
entries such that all entries with the same weight $\rat_i$ are in the same
group. Then, if we obtain only one group, the RV $\denV$ will be gamma distributed
with shape $\nMem_1$ and scale $\rat_1$
\begin{equation}
  \label{eq:pdfBfY1}
  p_\denV ( \denV ) = \frac{1}{\Gamma (\nMem_1) \rat_1^{\nMem_1}}
    \denV^{\nMem_1-1} e^{-\frac{\denV}{\rat_1}}.
\end{equation}
If we get $\nGrp \!>\! 1$ then the PDF of $\denV$ can be expressed according to
\cite{CuZhFe2004} as
\begin{equation}
  \label{eq:pdfBfY}
  p_\denV ( \denV ) = \sum_{i=1}^\nGrp \sum_{j=1}^{\nMem_i}
    \expC_{ij} \frac{1}{\Gamma (j) \rat_i^j} \denV^{j-1}
    e^{-\frac{\denV}{\rat_i}},
\end{equation}
where the coefficients $\expC_{ij}$ are
\begin{equation}
  \expC_{ij} \!=\! (-1)^{\nMem_i \!+\! j} \sum_{\tuple (i,j)} \prod_{ \substack{
    k = 1 \\ k \neq i } }^\nGrp \binom{\nMem_k \!+\! q_k \!-\! 1}{q_k}
    \frac{\left( \frac{\rat_k}{\rat_i} \right)^{q_k}}
    {\left( 1 \!-\! \frac{\rat_k}{\rat_i} \right)^{\nMem_k \!+\! q_k}},
\end{equation}
where $\tuple (i,j)$ is a set of $\nGrp$-tuples with nonnegative integers according
to
\begin{equation}
  \tuple (i,j) = \bigg\{ \! \left( q_1 ~ q_2 ~ \cdots ~ q_\nGrp \right) : q_i \!=\!
    0,~ \sum_{k=1}^\nGrp q_k = \nMem_i \!-\! j \bigg\}.
\end{equation}
Distributions of $\numV$ and $\denV$ and may now
be used in \eqref{eq:sinrPdf} to derive the probability of outage. With
$\nGrp \!=\! 1$, we use \eqref{eq:pdfBfX} and \eqref{eq:pdfBfY1}
that leads to
\addtocounter{equation}{3}
\begin{align}
  f_\sinr (\sinr)
  &=
    \int_0^\infty
    (\denV \!+\! 1)
    \sum_{k=1}^\minL
    \sum_{l=\maxL-\minL}^{(\maxL \!+\! \minL \!-\! 2k)k}
    \eigC_{kl}
    \frac{(\denV \!+\! 1)^l \sinr ^l}{l!}
    \left( \frac{k}{\ratOwn} \right)^{l+1} \nonumber \\
  &\quad
    \times e^{-\frac{(\denV+1)\sinr k}{\ratOwn}}
    \frac{\denV^{\nMem_1-1}}{\Gamma (\nMem_1)}
    \left( \frac{1}{\rat_1} \right)^{\nMem_1}
    e^{-\frac{\denV}{\rat_1}} d\denV \\
  &=
    \sum_{k=1}^\minL
    \sum_{l=\maxL-\minL}^{(\maxL \!+\! \minL \!-\! 2k)k}
    \frac{\eigC_{kl}}{l! \Gamma (\nMem_1)}
    \left( \frac{k}{\ratOwn} \right)^{l+1}
    \left( \frac{1}{\rat_1} \right)^{\nMem_1}
    \sinr ^l
    e^{-\frac{k \sinr}{\ratOwn}} \nonumber \\
  &\quad
    \times \int_0^\infty
    (\denV \!+\! 1)^{l+1}
    \denV^{\nMem_1-1}
    e^
      { -\denV \left(
        \frac{k \sinr}{\ratOwn} + \frac{1}{\rat_1}
      \right) }
    dy.
\end{align}
Now, by first applying \cite[(1.111)]{GrRy2007}
before using \cite[(3.351.3)]{GrRy2007}
we can derive the PDF of SINR as
\begin{multline}
  f_\sinr (\sinr)
  =
    \sum_{k=1}^\minL
    \sum_{l=\maxL-\minL}^{(\maxL \!+\! \minL \!-\! 2k)k}
    \eigC_{kl}
    \sinr ^l
    e^{-\frac{k \sinr}{\ratOwn}}
    \sum_{r=0}^{l+1}
    \binom{l \!+\! 1}{r}
    \frac{\Gamma (r \!+\! \nMem_1)}{l! \Gamma (\nMem_1)} \\
  \quad
    \times 
    \left( \frac{k}{\ratOwn} \right)^{l+1}
    \left( \frac{1}{\rat_1} \right)^{\nMem_1}
    \left(
      \frac{k \sinr}{\ratOwn} + \frac{1}{\rat_1}
    \right)^{-(r+\nMem_1)}.
\end{multline}
For a general case with $\nGrp \!>\! 1$ the PDF has been derived in \cite{Ah2009}
and is given in \eqref{eq:sinrPdfBf}, with $\sOne_i \!=\! \ratOwn / \rat_i$. In a
similar way we may use \eqref{eq:pdfBfX} and \eqref{eq:pdfBfY1} or \eqref{eq:pdfBfY}
in \eqref{eq:pOut} to calculate the outage probability. With $\nGrp \!=\! 1$ we
get
\begin{align}
  \pOut \!
  &= \! \int_0^{\sinr_0} \!\!\! \int_0^\infty \!\! (\denV \!+\! 1) \sum_{k=1}^\minL \!
    \sum_{l=\maxL-\minL}^{(\maxL+\minL-2k)k} \!\!\! \eigC_{kl} \frac{(\denV \!+\! 1)^l
    \sinr ^l}{\Gamma (l \!+\! 1)} \! \left( \frac{k}{\ratOwn} 
    \right)^{l+1} \nonumber \\
  &\quad \! \times e^{-\frac{(\denV+1)\sinr k}{\ratOwn}} \frac{\denV^{\nMem_1-1}}
    {\Gamma (\nMem_1)} \left( \frac{1}{\rat_1} \right)^{\nMem_1}
    e^{-\frac{\denV}{\rat_1}} d\denV d\sinr \\
  &= \! \sum_{k=1}^\minL \! \sum_{l=\maxL-\minL}^{(\maxL+\minL-2k)k} \!\!\!
    \frac{\eigC_{kl}}
    {\Gamma (l \!+\! 1) \Gamma (\nMem_1)} \left( \frac{k}{\ratOwn} \right)^{l+1} \!
    \left( \frac{1}{\rat_1} \right)^{\nMem_1} \! \nonumber \\
  &\quad \! \times \!\! \int_0^\infty \!\! (\denV \!+\! 1)^{l+1} \denV^{\nMem_1-1}
    e^{-\frac{\denV}{\rat_1}} \!\! \int_0^{\sinr_0} \!\! \sinr^l e^{-\frac{(\denV+1)
    \sinr k}{\ratOwn}} d\sinr d\denV
\end{align}
We now use \cite[(3.351.1)]{GrRy2007} and proceed
\begin{align}
  \pOut \!
  &= \! \sum_{k=1}^\minL \! \sum_{l=\maxL-\minL}^{(\maxL+\minL-2k)k} 
    \frac{\eigC_{kl}}{\Gamma (\nMem_1)} \left( \frac{1}{\rat_1} \right)^{\nMem_1} \!
    \int_0^\infty \denV^{\nMem_1-1} e^{-\frac{\denV}{\rat_1}} \nonumber \\
  &\quad \! \times \! \left( \! 1 \!-\! e^{\frac{-(\denV \!+\! 1) \sinr_0 k}
    {\ratOwn}} \! \sum_{r=0}^l \! \frac{1}{r!} \! \left(
    \frac{(\denV \!+\! 1) \sinr_0 k}{\ratOwn} \right)^r \right) d\denV.
\end{align}
Applying \cite[(1.111)]{GrRy2007} and \cite[(3.351.3)]{GrRy2007} we get the final form
\eqref{eq:pOutBf1}. Outage probability of the case with $\nGrp \!>\! 1$ has
been derived in a similar way in \cite{Ah2009}, we show it in \eqref{eq:pOutBf}.

\subsection{Outage Probability of OSTBC}
\label{ssec:OSTBC}

We will start the analysis of OSTBC for $2 \!\times\! 2$ MIMO case and
subsequently generalize it for higher dimensions. Let us denote
the received signal before filtering by
\begin{equation}
  \rV = \rVown + \sum_{i=1}^\nInt \rVint_i + \nV,
\end{equation}
where $\rVown$ denotes the useful signal part and $\rVint_i$ denotes the
interference part from $i$-th iBS. The useful part of the received signal may be
expressed as
\begin{equation}
  \label{eq:ostbcOwnRvector}
  \left[
    \begin{array}{l}
      \rElOwn_1^{(1)} \\ \rElOwn_1^{(2)\cc} \\
	  \rElOwn_2^{(1)} \\ \rElOwn_2^{(2)\cc}
    \end{array} 
  \right] =
  \sqrt{\rPown}
  \left[
    \begin{array}{rr}
      \fElOwn_{11} & \fElOwn_{12} \\ \fElOwn_{12}^\cc & -\fElOwn_{11}^\cc \\
      \fElOwn_{21} & \fElOwn_{22} \\ \fElOwn_{22}^\cc & -\fElOwn_{21}^\cc
    \end{array}
  \right]
  \left[
    \begin{array}{c}
      \dEown^{(1)} \\ \dEown^{(2)}
    \end{array}
  \right],
\end{equation}
where $m$ in $\rElOwn_m^{(n)}$ represents receive antenna index, $n$ in
$\rElOwn_m^{(n)}$ represents time instance/symbol index, $\fElOwn_{mn}$ is an
element of $\ffOwn$, $m$ in $\dEown^{(m)}$ represents time instance index and
$^\cc$ denotes complex conjugate. If $j$-th interferer also performs OSTBC,
the vector $\rVint_j$ will have the same structure as
\eqref{eq:ostbcOwnRvector} with correspondingly different channel and symbol
values. For $k$-th interferer performing beamforming, omitting the $k$ index
where it reduces readability, the received signal will be
\begin{equation}
  \left[
    \begin{array}{c}
      \rElInt_1^{(1)} \\ \rElInt_1^{(2)} \\ \rElInt_2^{(1)} \\ \rElInt_2^{(2)}
    \end{array}
  \right]_k =
  \sqrt{\rP_k}
  \left[
    \begin{array}{c}
      \dE^{(1)} \left( \fElInt_{11} \pE_1 + \fElInt_{12} \pE_2 \right) \\
      \dE^{(2)} \left( \fElInt_{11} \pE_1 + \fElInt_{12} \pE_2 \right) \\
      \dE^{(1)} \left( \fElInt_{21} \pE_1 + \fElInt_{22} \pE_2 \right) \\
      \dE^{(2)} \left( \fElInt_{21} \pE_1 + \fElInt_{22} \pE_2 \right)
    \end{array}
  \right],
\end{equation}
where $\fElInt_{mn}$ denotes element of $\ff_k$ and $\pE_m$ denotes element of
$\pV_k$. For $l$-th interferer performing spatial multiplexing, omitting the
$l$ index where it reduces readability, we get
\begin{displaymath}
  \!
  \left[ \!\!\!
    \begin{array}{c}
      \rElInt_1^{(1)} \\ \rElInt_1^{(2)} \\ \rElInt_2^{(1)} \\ \rElInt_2^{(2)}
    \end{array}
  \!\!\! \right]_l \!\!\!\!\! = \!\!
  \sqrt{\rP_l} \!\!
  \left[ \!\!\!
    \begin{array}{c}
      \dE_1^{(1)} \!\!
        \left( \fElInt_{11} \pE_{11} \!+\! \fElInt_{12} \pE_{21} \right) \!+\!
      \dE_2^{(1)} \!\!
        \left( \fElInt_{11} \pE_{12} \!+\! \fElInt_{12} \pE_{22} \right) \\
      \dE_1^{(2)} \!\!
        \left( \fElInt_{11} \pE_{11} \!+\! \fElInt_{12} \pE_{21} \right) \!+\!
      \dE_2^{(2)} \!\!
        \left( \fElInt_{11} \pE_{12} \!+\! \fElInt_{12} \pE_{22} \right) \\
      \dE_1^{(1)} \!\!
        \left( \fElInt_{21} \pE_{11} \!+\! \fElInt_{22} \pE_{21} \right) \!+\!
      \dE_2^{(1)} \!\!
        \left( \fElInt_{21} \pE_{12} \!+\! \fElInt_{22} \pE_{22} \right) \\
      \dE_1^{(2)} \!\!
        \left( \fElInt_{21} \pE_{11} \!+\! \fElInt_{22} \pE_{21} \right) \!+\!
      \dE_2^{(2)} \!\!
        \left( \fElInt_{21} \pE_{12} \!+\! \fElInt_{22} \pE_{22} \right)
    \end{array}
  \!\!\! \right] \!\!,
\end{displaymath}
where $m$ in $\dE_m^{(n)}$ represents transmission layer index and $\pE_{mn}$
is an element of $\pM_l$. We get our symbol estimates $\sEst$ from
$\sEst \!=\! \rF \rV$ where $\rF$ is the receive filter
\begin{equation}
  \rF =
  \left[
    \begin{array}{rrrr}
      \fElOwn_{11}^\star & \fElOwn_{12} & \fElOwn_{21}^\star & \fElOwn_{22} \\
      \fElOwn_{12}^\star & -\fElOwn_{11} & \fElOwn_{22}^\star & -\fElOwn_{21}
    \end{array}
  \right].
\end{equation}

\begin{figure}[t]
  \begin{center}
    \includegraphics[scale=0.5]{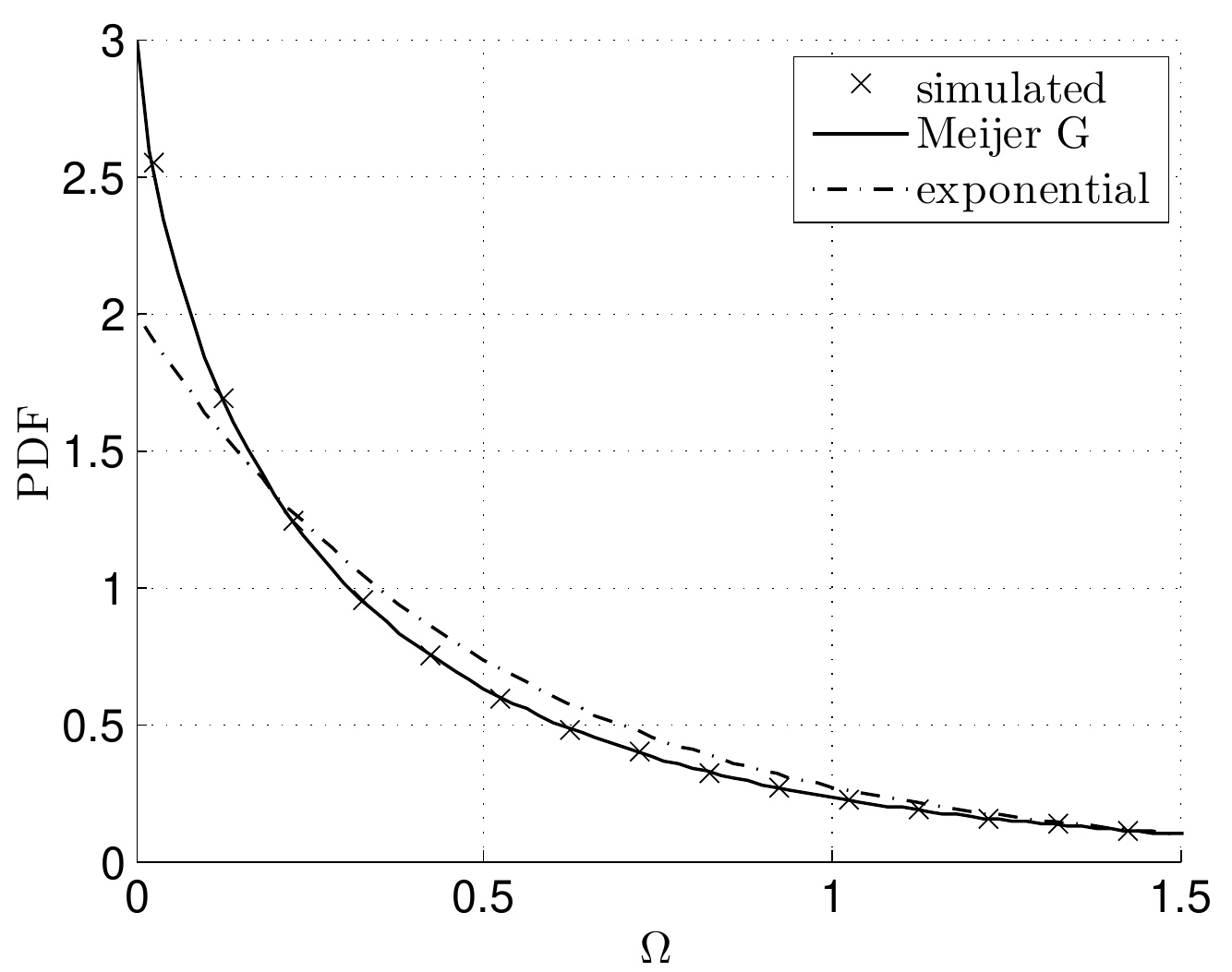}
    \caption{Approximation of $\stbcSone_m$ term for outage probability of
      OSTBC transmission in $2 \!\times\! 2$ MIMO case with the
      interferer performing spatial multiplexing. Meijer G assumes
      $\stbcStwo_m$ and $\stbcSthree_m$ to be independent, exponential PDF is
      our final approximation.}\label{fig:stbcApprox}
  \end{center}
\end{figure}

\noindent The numerator RV $\numV$ of \eqref{eq:sinrFrac} is known
\cite{RoBoPa2005} to be
\begin{equation}
  \label{eq:stbcX}
  \numV = \frac{\rPown}{4 \nP} \left\Vert \ffOwn \right\Vert _\text{F} ^2,
\end{equation}
where $\left\Vert \ffOwn \right\Vert _\text{F}$ is a Frobenius norm of $\ffOwn$.
Note that the quadruple noise power in the denominator of \eqref{eq:stbcX} comes
from 1) transmission power normalization and 2) processing noise samples from
two time instances at once \cite{ChHiTaHo2007}. The numerator $\numV$ is hence
gamma distributed with shape $\nR \nT$ and scale $\ratOwn \!=\! \rPown / \nT^2
\nP$.

\begin{figure*}[!b]
  \hrulefill
  \vspace*{0pt}
  \normalsize
  \setcounter{MYtempeqncnt}{\value{equation}}
  \setcounter{equation}{41}
  \begin{equation}
    \label{eq:sinrPdfStbc}
    f_\sinr (\sinr) \approx
    \sum_{i=1}^\nGrp
    \sum_{j=1}^{\nMem_i}
    \expC_{ij}
    \sinr^{\nR \nT - 1}
    e^{-\frac{\sinr}{\ratOwn}}
    \left( \frac{1}{\ratOwn} \right)^{\nR \nT}
    \left( \frac{1}{\rat_1} \right)^j
    \sum_{r=0}^{\nR \nT}
    \binom{\nR \nT}{r}
    \frac{\Gamma (r + j)}{\Gamma (\nR \nT) \Gamma (j)}
    \left(
      \frac{\sinr}{\ratOwn} + \frac{1}{\rat_1}
    \right)^{-(r + j)}
  \end{equation}
  \begin{equation}
    \label{eq:pOutStbc}
    \pOut \approx
    \sum_{i=1}^\nGrp
    \sum_{j=1}^{\nMem_i}
    \expC_{ij}
    \left(
      1 -
      e^{-\frac{\sinr_0}{\rat_0}}
      \left( \frac{1}{\rat_i} \right)^j
      \sum_{r=0}^{\nR \nT - 1}
      \sum_{s=0}^r
      \binom{r}{s}
      \frac{\Gamma (j+s)}{r! \Gamma (j)}
      \left( \frac{\sinr_0}{\rat_0} \right)^r
      \left(
        \frac{\sinr_0}{\rat_0} + \frac{1}{\rat_i}
      \right)^{-(j+s)}
    \right)
  \end{equation}
  \setcounter{equation}{\value{MYtempeqncnt}}
\end{figure*}

The denominator RV $\denV$ is a sum of RVs $\denV_i$, the shape of which is
determined by the MIMO transmission technique of the interferers. If $j$-th
interferer performs OSTBC \cite{ChHiTaHo2007}, its contribution $\denV_j$ is
given by a sum of $\nT$ exponentially distributed RVs with rate $1 / \rat_j
\!=\! \nT^2 \nP / \rP_j$. For $k$-th interferer performing beamforming we can write
\begin{equation}
  \denV_k =
  \frac
    {\rP_k}
    {2 \nP}
  \left(
    \stbcSone _1 + \stbcSone _2
  \right),
\end{equation}
where $\stbcSone_m$ represent independent power contribution from $m$-th
time instance/transmission symbol. These contributions, omitting the $k$ index
where it reduces readability, are
\begin{displaymath}
  \stbcSone _1 \!=\!
  \frac
    {
      \left\vert
        \fElOwn_{11}^\cc \dE^{(1)} \!\!
        \left(
          \fElInt_{11} \pE_1 \!\!+\! \fElInt_{12} \pE_2
        \right) \!+\!
        \fElOwn_{21}^\cc \dE^{(1)} \!\!
        \left(
          \fElInt_{21} \pE_1 \!\!+\! \fElInt_{22} \pE_2
        \right)
      \right\vert ^2
	}
    {\left\Vert \ffOwn \right\Vert _\text{F} ^2},
\end{displaymath}
\begin{displaymath}
  \stbcSone _2 \!=\!
  \frac
    {
      \left\vert
        \fElOwn_{12} \dE^{(2)} \!\!
        \left(
          \fElInt_{11} \pE_1 \!\!+\! \fElInt_{12} \pE_2
        \right) \!+\!
        \fElOwn_{22} \dE^{(2)} \!\!
        \left(
          \fElInt_{21} \pE_1 \!\!+\! \fElInt_{22} \pE_2
        \right)
      \right\vert ^2
    }
    {\left\Vert \ffOwn \right\Vert _\text{F} ^2}.
\end{displaymath}
As the data symbols and beamforming vectors are normalized to unity,
$\dE^{(m)} ( \fElInt_{n1} \pE_1 \!+\! \fElInt_{n2} \pE_2 )$ within
numerator of $\stbcSone_m$ remain complex Gaussian distributed with
zero mean and unit variance. However, because multiplication with elements of
$\ffOwn$ is not properly normalized, i.e., the numerator of $\stbcSone_m$
does not contain all elements of $\ffOwn$ that are present in the denominator,
the Gaussianity is lost and the distribution of $\stbcSone_m$ is not
straightforward to establish.

We thus propose an approximation. Illustrating our approach on
$\stbcSone_1$, let us write $\stbcSone_1 \!=\! \stbcStwo_1 \stbcSthree_1$,
where
\begin{displaymath}
  \stbcStwo _1 \!=\!
  \frac
    {
      \left\vert
        \fElOwn_{11}^\cc \dE^{(1)} \!\!
        \left(
          \fElInt_{11} \pE_1 \!\!+\! \fElInt_{12} \pE_2
        \right) \!+\!
        \fElOwn_{21}^\cc \dE^{(1)} \!\!
        \left(
          \fElInt_{21} \pE_1 \!\!+\! \fElInt_{22} \pE_2
        \right)
      \right\vert ^2
	}
    {
      \left\vert \fElOwn_{11} \right\vert ^2 +
      \left\vert \fElOwn_{21} \right\vert ^2
    },
\end{displaymath}
\begin{equation}
  \stbcSthree _1 =
  \frac
    {
      \left\vert
        \fElOwn_{11}
      \right\vert ^2 +
      \left\vert
        \fElOwn_{21}
      \right\vert ^2
    }
    {
      \left\Vert \ffOwn \right\Vert _\text{F} ^2
    }.
\end{equation}
The first RV $\stbcStwo_1$ is now properly normalized and follows exponential
distribution with rate $\nL ^{(k)}$. The second RV $\stbcSthree_1$ can be expressed as
$X / (X \!+\! Y)$ where $X$ is gamma distributed with rate $\nR$ and unit scale
and $Y$ is gamma distributed with rate $\nR (\nT \!-\! 1)$ and unit scale. Variable
$\stbcSthree_1$
therefore follows beta distribution with shape parameters $\alpha \!=\! \nR$
and $\beta \!=\! \nR (\nT \!-\! 1)$.

Variables $\stbcStwo_m$ and $\stbcSthree_m$ are generally not independent. For the
sake of tractability we will therefore make our first approximation step and assume
them to be
independent. Variable $\stbcSone_m$ is thus given by a product of independent
exponential and beta RVs. Because exponential distribution is a special case of gamma
distribution, a PDF of such product is known \cite{SpTh1970} and in our case is
\begin{equation}
  \label{eq:meijerApprx}
  f_{\stbcSone_m} (x) \approx
  \frac{\nL^{(k)} \Gamma (\nR \nT)}{\Gamma (\nR)}
  G_{1,2}^{2,0} \!
  \left( \!\!
    \left. \!
      \begin{array}{c}
        \nT \nR \!-\! 1 \\ \nR \!-\! 1,0
      \end{array}
    \!\! \right\vert \! \nL^{(k)} x
  \! \right) \!,
\end{equation}
where $G_{p,q}^{m,n}$ is the Meijer G-function.  Using \cite[(7.811)]{GrRy2007}
that states
\begin{multline}
  \int _0^\infty x^{\rho -1} G_{p,q}^{m,n}
  \left( \!
    \left.
      \begin{array}{c}
        a_1,\dots ,a_p \\ b_1,\dots ,b_q
      \end{array}
    \right\vert \alpha x
  \! \right) dx \\ =
  \frac
    {
      \prod _{j=1}^m \Gamma (b_j \!+\! \rho )
      \prod _{j=1}^n \Gamma (1 \!-\! a_j \!-\! \rho )
    }
    {
      \prod _{j=m+1}^q \Gamma (1 \!-\! b_j \!-\! \rho )
      \prod _{j=n+1}^p \Gamma (a_j \!+\! \rho )
    }
  \alpha ^{-\rho }
\end{multline}
we can derive the mean value of $\stbcSone_m$ to be $1 / \nT \nL^{(k)}$. This
approximate PDF of $\stbcSone_m$ is not exactly convenient to work with. However,
we look at its shape, as e.g. in \figurename~\ref{fig:stbcApprox}, and realize it
is remarkably close to that of an exponential distribution. Hence, as the
second step of our approximation we assume $\stbcSone_m$ to take on a shape of
exponential RV with rate $\nT \nL^{(k)}$. We note here that while this final
shape is the same as originally proposed in \cite{LiCiHi2008} and
subsequently used in \cite{LiZhCiZh2011_2,ZhLiCi2012}, our intermediate
approximation \eqref{eq:meijerApprx} is novel and more precise. It also
illustrates the way one comes up with the final approximation using
exponential distribution in a more insightful manner.

Hence, $k$-th iBS, whether it performs precoding or OSTBC, contributes to
$\denV$ by a sum of $\nT \nL^{(k)}$ terms $\stbcSone_m$. Each of the $\stbcSone_m$
terms is exponentially distributed with rate $1 / \rat_k \!=\! \nT^2
\nL^{(k)} \nP / \rP_k$, in case of OSTBC exactly and in case of precoding
approximately. Now, as $\denV$ is again given by a sum of weighted
exponentially distributed RVs, we can divide them in $\nGrp$ groups such
that $i$-th group collects all $\nMem_i$ contributions that have the same
weight $\rat_i$. Because this is the same case is in Subsection
\ref{ssec:BF}, the formulas for PDF of $\denV$ in \eqref{eq:pdfBfY1} and
\eqref{eq:pdfBfY} are valid also when UE receives OSTBC transmission.

Having PDFs of $\numV$ and $\denV$ ready we can use them along
\cite[(1.111)]{GrRy2007} and \cite[(3.351.3)]{GrRy2007} in \eqref{eq:sinrPdf} to derive the
PDF of SINR. With $\nGrp \!=\! 1$, which is the case with equi-power
interference contributions, we get
\begin{multline}
  f_\sinr (\sinr)
  \approx
  \sinr^{\nR \nT -\! 1}
  e^{-\frac{\sinr}{\ratOwn}}
  \!\!
  \sum_{r=0}^{\nR \nT}
  \!\!
  \binom{\nR \nT}{r}
  \left( \frac{1}{\ratOwn} \right)^{\nR \nT} \\
  \times
  \!
  \left( \frac{1}{\rat_1} \right)^{\nMem_1}
  \!\!
  \frac{\Gamma (r \!+\! \nMem_1)}{\Gamma (\nR \nT) \Gamma (\nMem_1)}
  \!
  \left(
    \frac{\sinr}{\ratOwn} \!+\! \frac{1}{\rat_1}
  \right)^{-(r \!+\! \nMem_1)}.
\end{multline}
For a general case with $\nGrp \!>\! 1$ the formula is given in \eqref{eq:sinrPdfStbc}.
Using $\numV$ and $\denV$ along with \cite[(3.351.1)]{GrRy2007},\cite[(1.111)]{GrRy2007} and
\cite[(3.351.3)]{GrRy2007} we can also derive the outage probability. With
$\nGrp \!=\! 1$ we get
\begin{multline}
  \label{eq:pOutStbc1}
  \pOut \approx 1 -
  e^{- \frac{\sinr_0}{\rat_0}}
  \left( \frac{1}{\rat_1} \right) ^{\nMem_1}
  \sum_{r=0}^{\nR \nT - 1}
  \sum_{s=0}^r
  \binom{r}{s}
  \frac{\Gamma (\nMem_1 + s)}{r! \Gamma (\nMem_1)} \\
  \times
  \left( \frac{\sinr_0}{\rat_0} \right)^r
  \left(
    \frac{\sinr_0}{\rat_0} + \frac{1}{\rat_1}
  \right)^{- (\nMem_1 + s)}
\end{multline}
and for the general case with $\nGrp \!>\! 1$ the result is given in
\eqref{eq:pOutStbc}.

\begin{figure}[t]
  \begin{center}
    \includegraphics[scale=0.5]{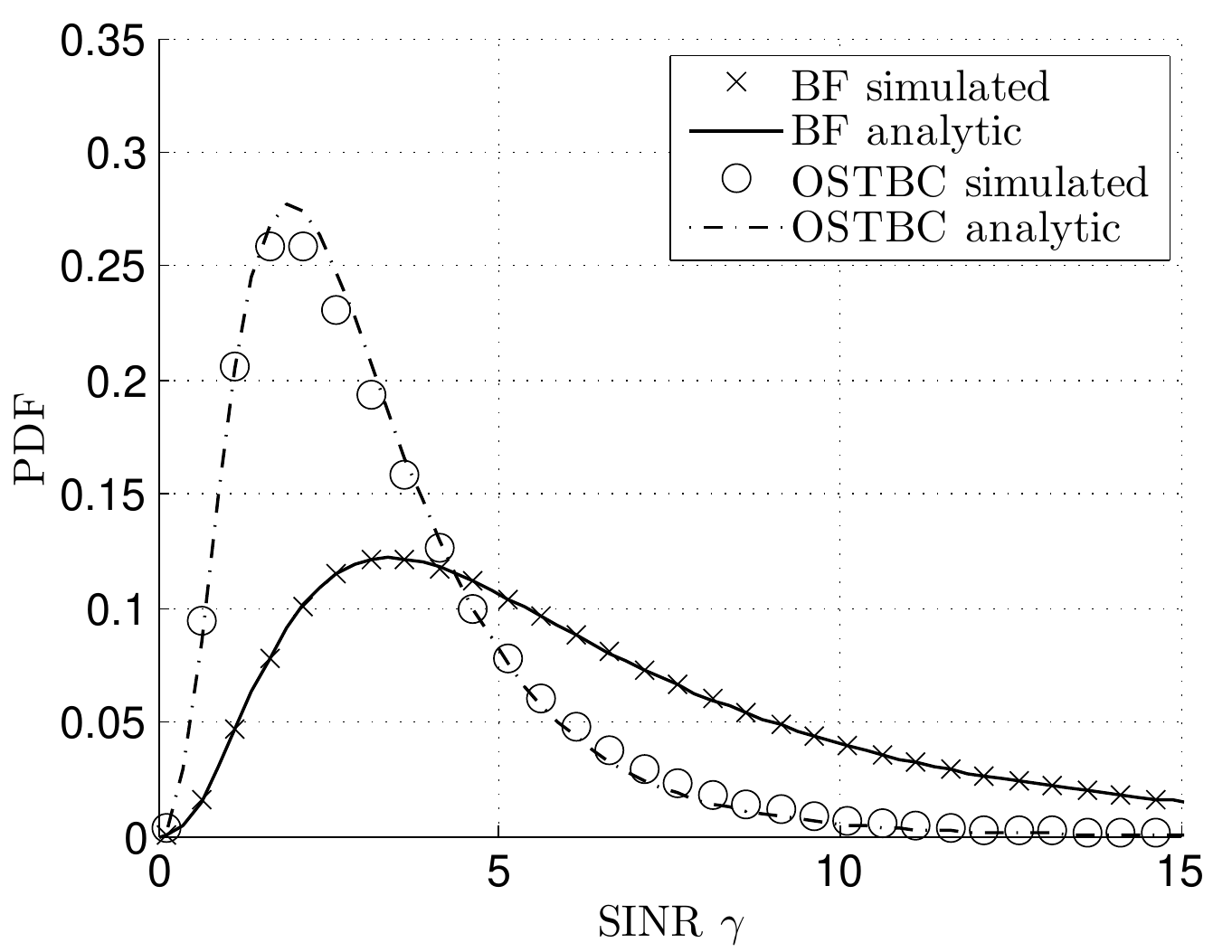}
    \caption{SINR distribution of beamforming and OSTBC under interference from
      unequi-power iBSs performing different multi-antenna transmission
      techniques.}\label{fig:sinrPrec}
  \end{center}
\end{figure}

\section{Results and Discussion}
\label{sec:resDis}

In this section we will verify the precision of our analysis from Section
\ref{sec:anl} by means of Monte Carlo simulations, present the main results on
the effect of interference rank on beamforming and OSTBC transmission and
discuss their significance and possible future work.

\subsection{Precision of the Analysis}

In order to demonstrate that our formulas of SINR distribution and probability
of outage are useful we perform Monte Carlo simulations and plot the collected
statistics side by side with outputs from the formulas. The statistics are
collected from $10^7$ independent channel realizations. Our reference input
parameters are:
\begin{itemize}
  \item Number of receive antennas $\nR \!=\! 2$.
  \item Number of transmit antennas $\nT \!=\! 2$.
  \item Noise power $\nP \!=\! 1$.
  \item $\mathrm{SNR} \!=\! 15\mathrm{dB}$. This
      corresponds to long term receive user signal power of $\rPown \!=\! 31.62$.
  \item Interference-to-noise-ratios $\mathrm{INR}_i \!=\! \left\{
      6\mathrm{dB}, 8\mathrm{dB}, 10\mathrm{dB}\right\}$,  corresponding to
      long term received interference powers of $\rP_i \!=\! \left\{ 3.98, 6.31, 10
      \right\}$, respectively.  The first iBS performs OSTBC, the second iBS performs
      beamforming, and the third iBS performs spatial multiplexing, respectively.
\end{itemize}
Firstly, we consider in \figurename~\ref{fig:sinrPrec} the PDF of SINR. A
mismatch between the analytical results and collected statistics may be seen at
two parts of the curve that corresponds to OSTBC
own transmission: the peak and the place where the right tail begins. This is due to
exponential distribution used to approximate \eqref{eq:meijerApprx} as shown in
\figurename~\ref{fig:stbcApprox}. Overall, the formulas for PDF of SINR show a good
match to the statistics collected from Monte Carlo simulations.

\begin{figure}[t]
  \begin{center}
    \includegraphics[scale=0.5]{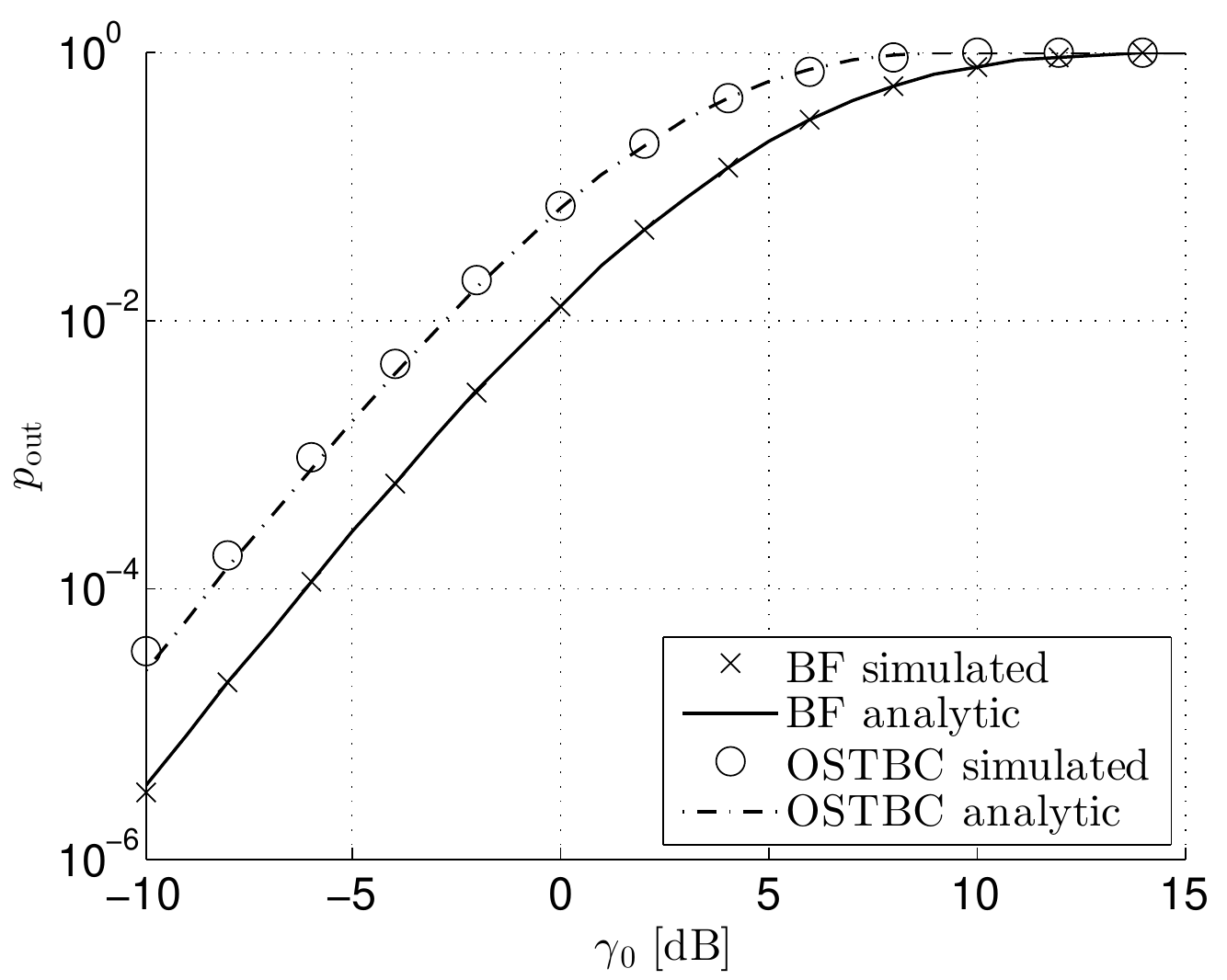}
    \caption{Outage probability of beamforming and OSTBC under interference from
      unequi-power iBSs performing different multi-antenna transmission
      techniques, as a function of outage threshold $\sinr_0$.}
  \label{fig:outProbPrec_varThres}
  \end{center}
\end{figure}

\begin{figure}[t]
  \begin{center}
    \includegraphics[scale=0.5]{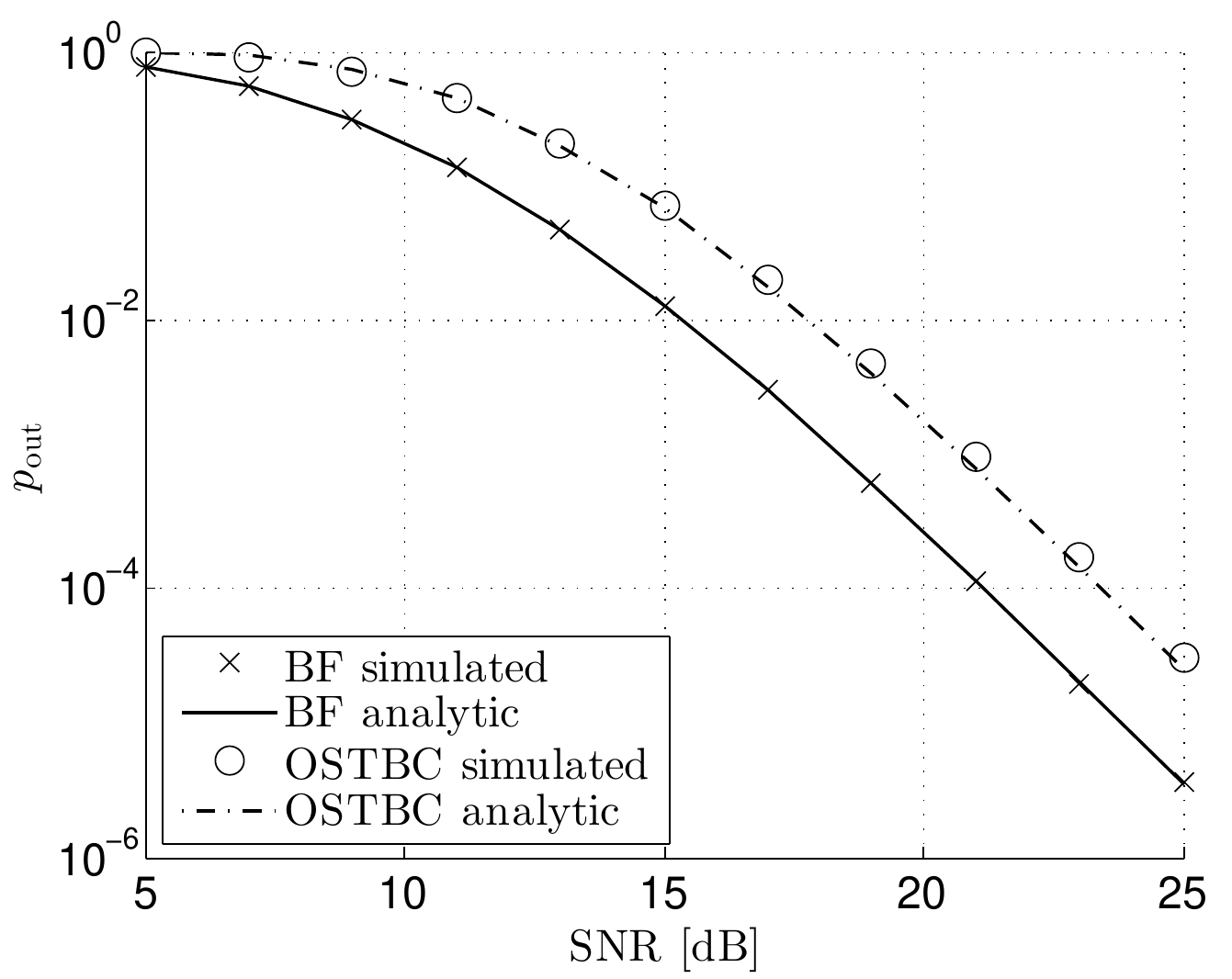}
    \caption{Outage probability of beamforming and OSTBC under interference from
      unequi-power iBSs performing different multi-antenna transmission
      techniques, as a function of signal-to-noise-ratio.}
    \label{fig:outProbPrec_varSNR}
  \end{center}
\end{figure}

Secondly we use Monte Carlo simulations
to corroborate the analytical results on outage probability derived in Section
\ref{sec:anl}. In \figurename~\ref{fig:outProbPrec_varThres} we plot probability
of outage as a function of $\sinr_0$ threshold whereas in
\figurename~\ref{fig:outProbPrec_varSNR} we provide the probability of outage as a function
of $\mathrm{SNR}$ with $\sinr_0 \!=\! 0\mathrm{dB}$.
Other parameters remain the same values as previously described.
Hence, the results in \figurename~\ref{fig:outProbPrec_varThres} and
\figurename~\ref{fig:outProbPrec_varSNR} directly correspond
to SINR distributions in \figurename~\ref{fig:sinrPrec}. 
Our results illustrate good match between the
analytical results and the simulation results.
We also notice the expected performance
difference between beamforming and OSTBC 
resulting from the presence/absence of CSI at the sBS
transmitter.

\subsection{Impact of Interference Rank}

As one of our main contributions, we use the results on probability of outage in
Section \ref{sec:anl} to study the effect of interference rank on beamforming and
OSTBC.

\begin{figure}[t]
  \begin{center}
    \includegraphics[scale=0.5]{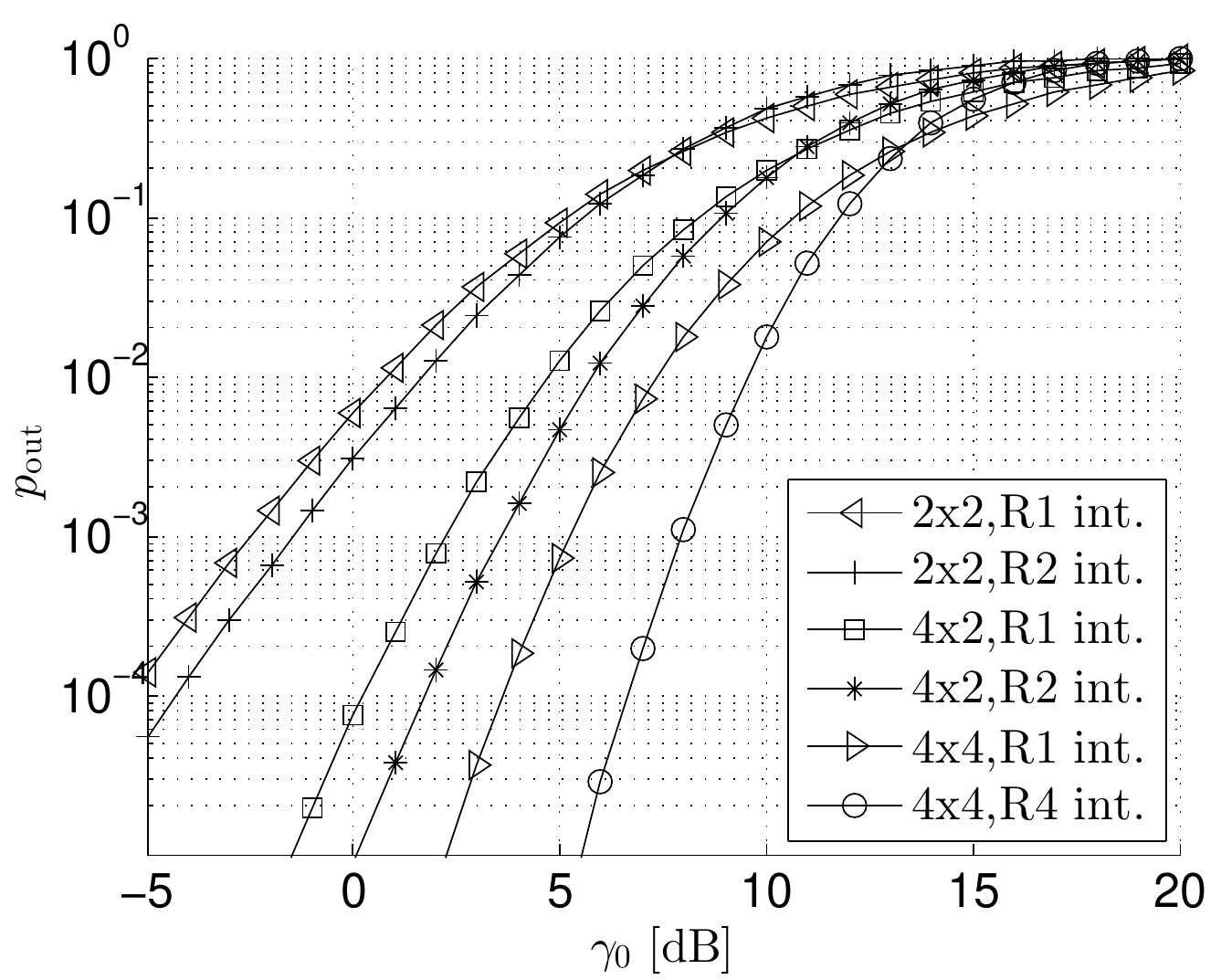}
    \caption{Outage probability of beamforming for different antenna configurations
      with a single iBS performing beamforming or spatial
      multiplexing. In the legend, R stands for rank, specifying
        interference rank.}
    \label{fig:bfOutProb}
  \end{center}
\end{figure}

A typical cellular user may be interfered from many iBSs that use various
multi-antenna techniques. However, only limited number of iBSs, called
dominant interferers, typically have a significant impact. These are most likely
iBSs that
are co-located with our UE of interest, or have a strong line-of-sight spatial
relation with it.
Furthermore, because limiting a transmission rank of an iBS may
have adverse effect on its own transmission, a single UE with weak link
should not limit performance of too many neighbors. For these reasons we will
draw our main insights from scenario with single iBS.  
We shall identify impact of multiple iBSs separately.

\begin{figure}[t]
  \begin{center}
    \includegraphics[scale=0.5]{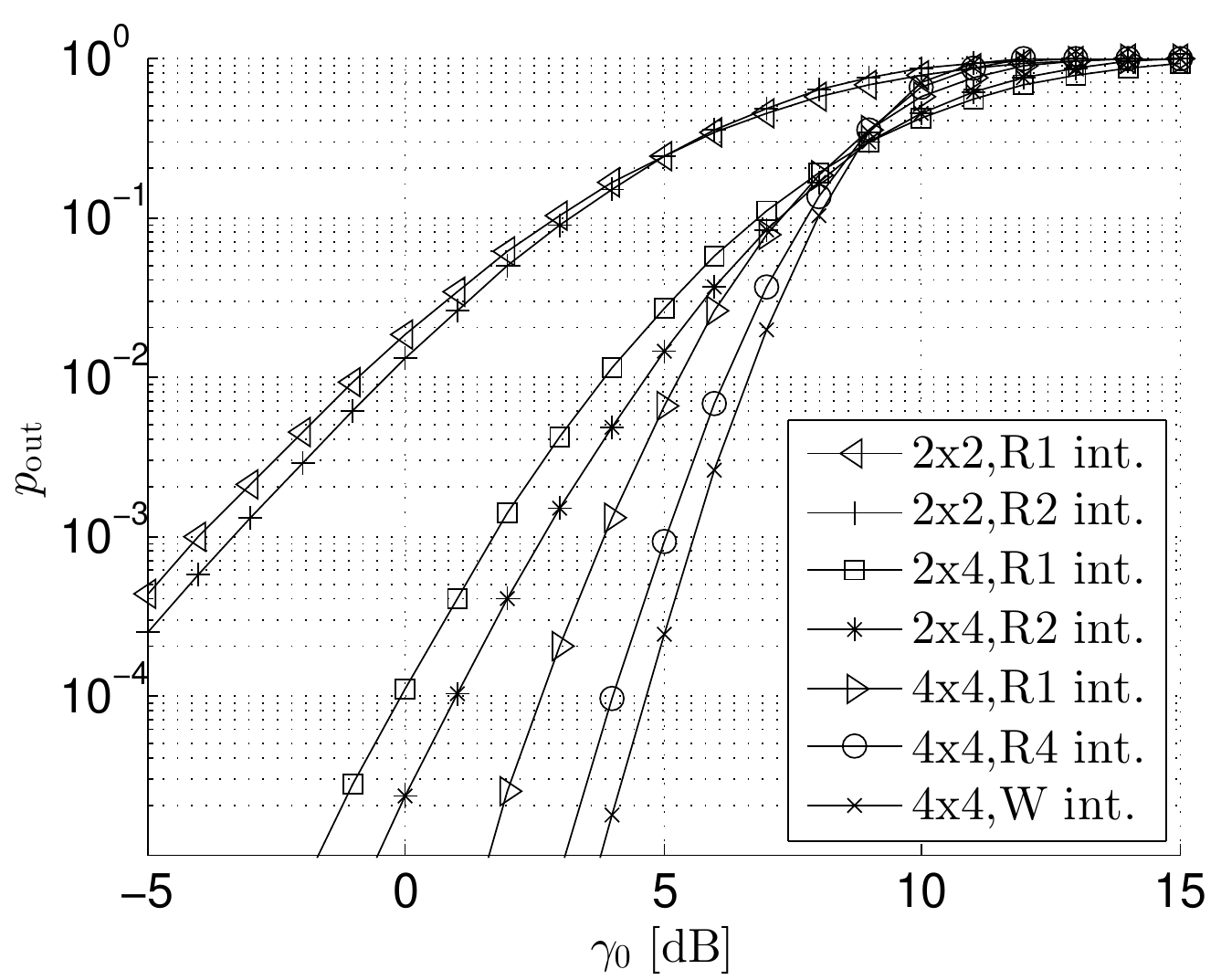}
    \caption{Outage probability of OSTBC for different antenna configurations
      with a single iBS performing beamforming or spatial
      multiplexing. In the legend, R stands for rank of the
        interferer and W int. denotes white interference.}\label{fig:stbcOutProb}
  \end{center}
\end{figure}

In \figurename~\ref{fig:bfOutProb} we show outage probability of beamforming with
a single iBS and $\mathrm{SNR} \!=\! 15\mathrm{dB}$
and $\mathrm{INR} \!=\! 10\mathrm{dB}$. The outage probability is plotted against
the $\sinr_0$ threshold, with different antenna configurations and different
interference rank. For every antenna configuration the two curves (rank 1
interference vs. higher rank interference) cross each other. Hence for
$\sinr_0$ above the crossing point, rank 1 interference causes lower probability of
outage while with $\sinr_0$ below the crossing point, rank 1 interference leads to
higher probability of outage. However, because such crossing points
correspond to
probability of outage well above $0.1$, we can make a general observation that in
the useful range of outage probability higher rank of the interference leads to
lower probability of outage. This is the most interesting result of our study. 
The
reason for this behavior stems from the fact that when an interferer transmits
with higher rank, it divides available power into weaker spatial streams and
lowers the possibility that much interference power may be directed towards the
UE of interest. In other words, spreading the interference into multiple spatial
streams leads to higher degrees of freedom in interference statistics, thereby
decreasing the probability of reducing the instantaneous SINR at the UE of interest.

\begin{figure}[t]
  \begin{center}
    \includegraphics[scale=0.5]{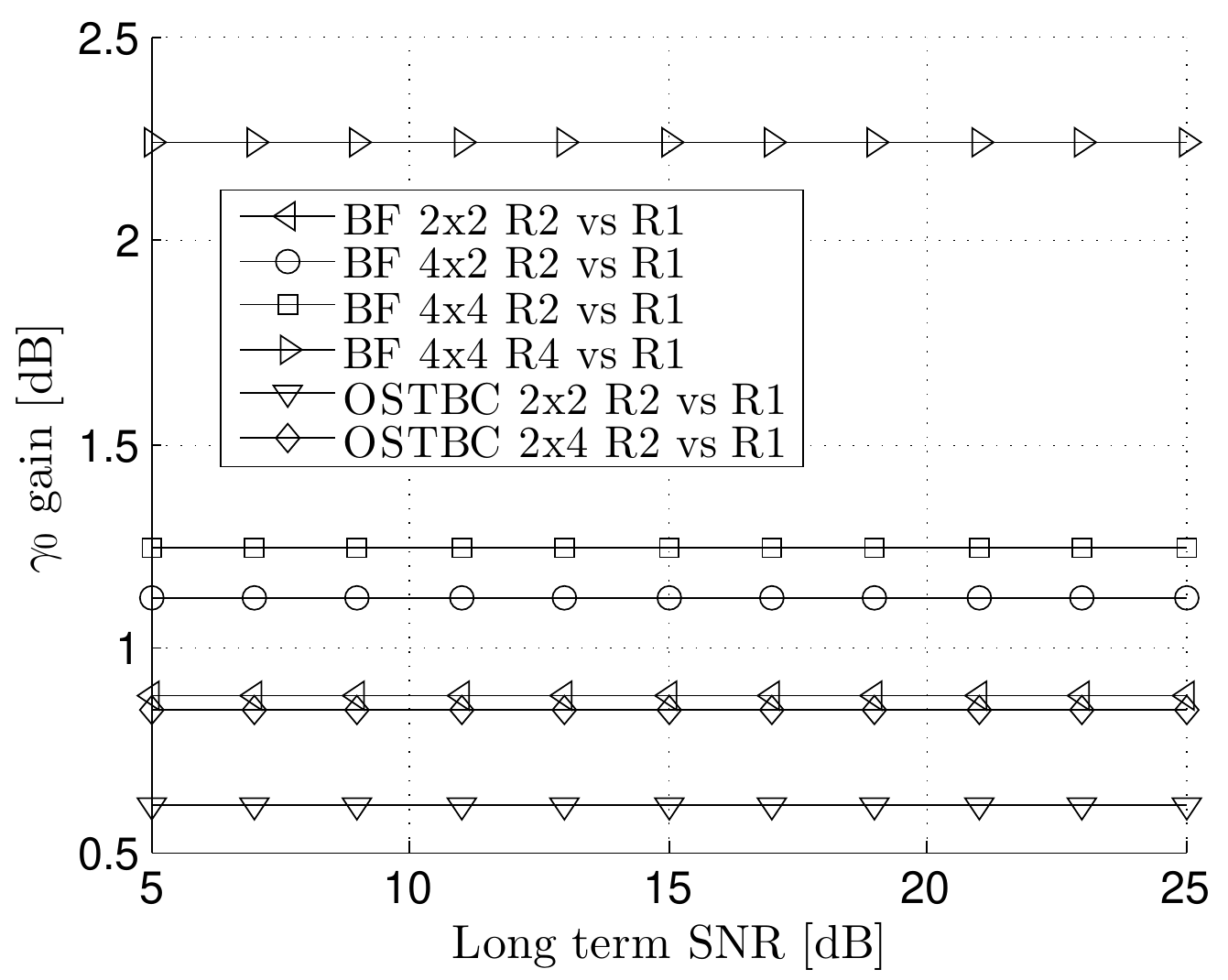}
    \caption{Gain in the supported outage threshold under single iBS with varying
      long term SNR and different antenna configurations. In the legend, R
      stands for rank of the
      interferer.}\label{fig:gainSNR}
  \end{center}
\end{figure}

For the own transmission using OSTBC we show results with the same parameter
settings in \figurename~\ref{fig:stbcOutProb}. Instead of $4 \!\times\! 2$ we
consider $2 \!\times\! 4$ MIMO configuration, $4 \!\times\! 4$ is included for
illustration purposes. Compared to sBS performing beamforming our observation
remains the same: more degrees of freedom in higher rank interference statistics
cause decrease in the probability of outage. However, the performance improvement,
i.e., the increase in supported $\sinr_0$ for a given $\pOut$ requirement, is not
as large with OSTBC as with beamforming. The reason may be found when comparing
the performance with the case when interference is white. In
\figurename~\ref{fig:stbcOutProb} we plot one such curve, outage probability of
$4 \!\times\! 4$ OSTBC transmission with white interference and the same
value of $\mathrm{INR} \!=\! 10\mathrm{dB}$. The highest possible
interference rank
brings the performance so close to the case with white interference that there is
only little space for further improvement.

\begin{figure}[t]
  \begin{center}
    \includegraphics[scale=0.5]{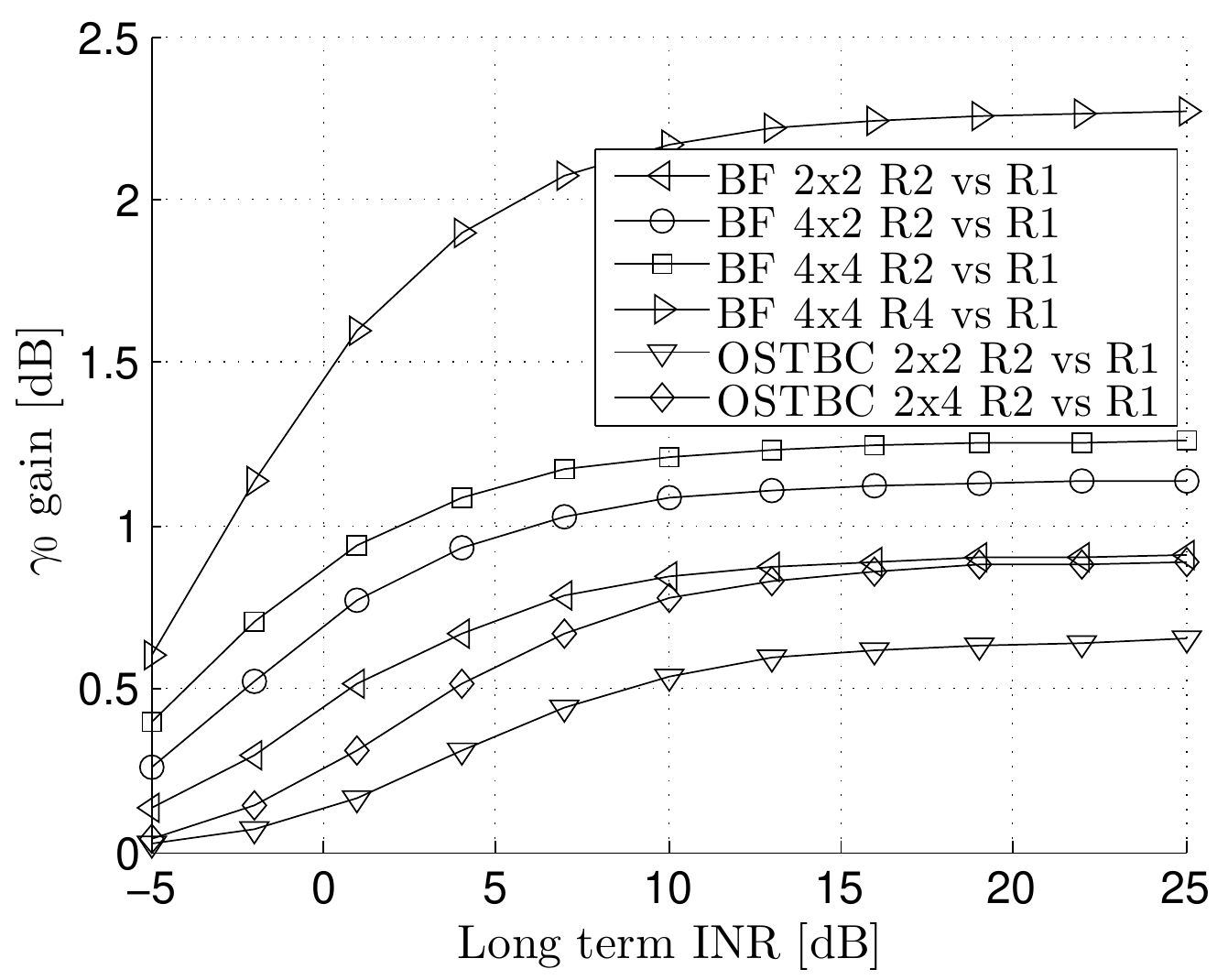}
    \caption{Gain in the supported outage threshold under single iBS with varying
      long term INR and different antenna configurations. In the legend, R
      stands for rank of the
      interferer.}\label{fig:gainINR}
  \end{center}
\end{figure}

Next, we studied the effect of SNR and INR on the results with
single iBS. As a performance
metric we used a $\sinr_0$ gain which we define as the increase in supported
outage threshold $\sinr_0$ at given outage requirement $\pOut \!=\! 0.01$.
Visually this represents a horizontal distance between curves corresponding
to rank 1 interference and higher rank interference in
\figurename~\ref{fig:bfOutProb} and \figurename~\ref{fig:stbcOutProb}.

In \figurename~\ref{fig:gainSNR} we show the $\sinr_0$ gain as
a function of long term SNR at $\mathrm{INR} \!=\! 15\mathrm{dB}$ for
different antenna configurations. The near constant SINR
means that $\sinr_0$ gain is indifferent to SNR. This is because when SNR
changes, SINR changes by the same amount, i.e., the curves of
outage probability versus $\sinr_0$ (\figurename~\ref{fig:bfOutProb} and
\ref{fig:stbcOutProb}) are only shifted along the horizontal axis.

In \figurename~\ref{fig:gainINR} we show the $\sinr_0$ gain
as a function of long term INR at $\mathrm{SNR} \!=\! 15\mathrm{dB}$
for different antenna configurations.
Here the $\sinr_0$ gain is an increasing function of INR. In lower INR range,
$\sinr_0$ increases faster, while in high INR range, the changes are less
significant. This is because when interference is much higher than noise,
changing interference power influences SINR almost as directly as changing
the  received user signal power (or SNR), which shifts the outage probability curve 
horizontally versus
$\sinr_0$. The $\sinr_0$ gain also noticeably
grows with higher number of transmit or receive antennas and with larger rank of
the iBS transmission. When sBS performs beamforming, the $\sinr_0$ gain
ranges from about $0.4\mathrm{dB}$ with $2 \!\times\! 2$ MIMO at $\mathrm{INR}
\!=\! 0\mathrm{dB}$ to more than $2\mathrm{dB}$ with $4 \!\times\! 4$ MIMO
at $\mathrm{INR} \!=\! 6\mathrm{dB}$ or higher. We consider the obtained
gains in case of beamforming worthwhile. On the other hand, with sBS performing
OSTBC the gains are relatively small, starting at around
$0.25\mathrm{dB}$
at $\mathrm{INR} \!=\! 0\mathrm{dB}$ but never exceeding $1\mathrm{dB}$.

\begin{figure}[t]
  \begin{center}
    \includegraphics[scale=0.5]{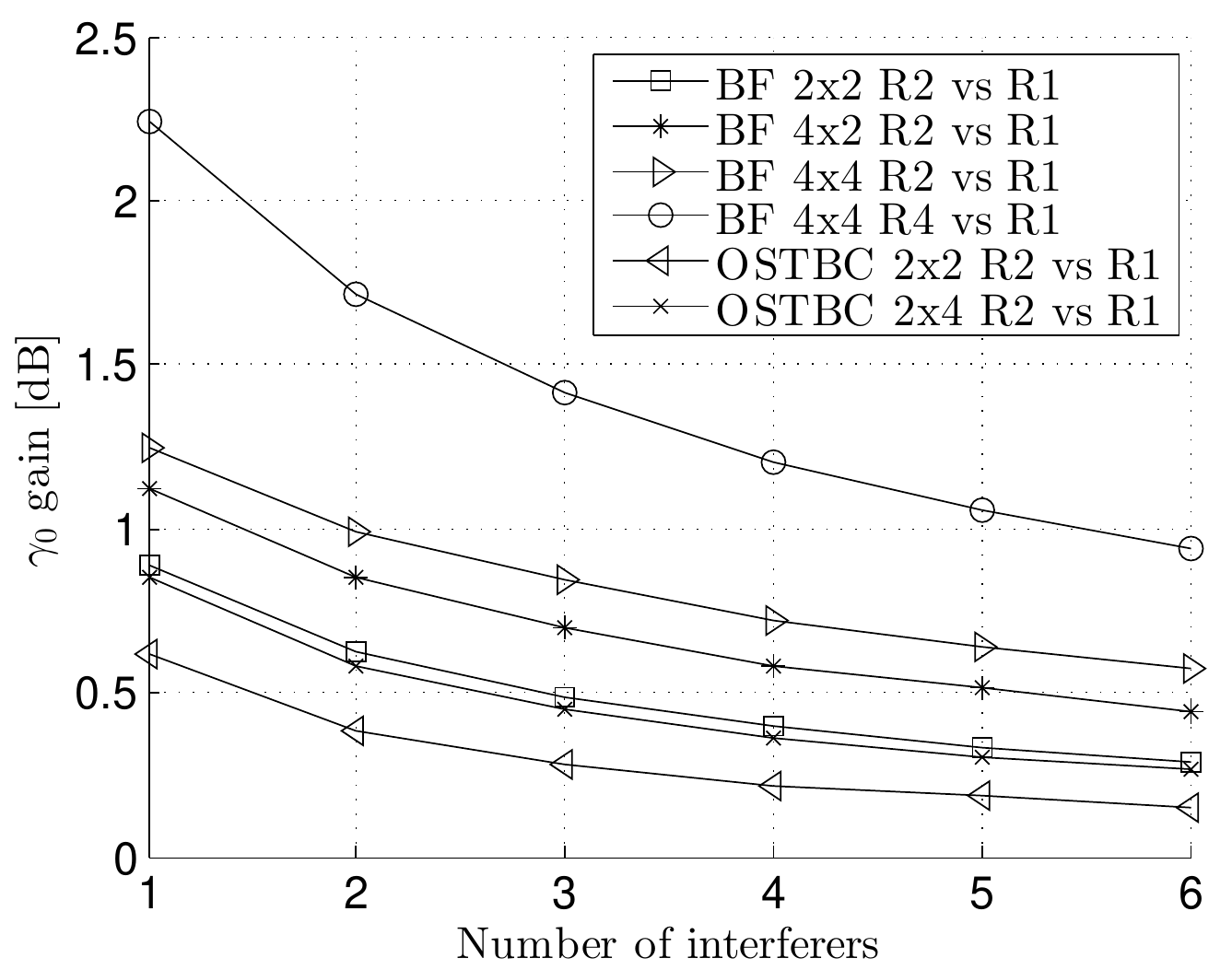}
    \caption{Gain in the supported outage threshold versus number of iBSs
      with constant sum interference power and different antenna
      configurations. In the legend, R stands for rank of the
      interferers.}\label{fig:gainNint}
  \end{center}
\end{figure}

Finally,  \figurename~\ref{fig:gainNint} shows the
$\sinr_0$ gain 
as a function of number of iBSs at $\mathrm{SNR} \!=\! 15\mathrm{dB}$. 
For specific number of iBSs the interferers
have equal transmission power, while across the number of iBSs we keep the total power of
interference at a constant value corresponding to $\mathrm{INR} \!=\!
15\mathrm{dB}$. The results show us that with more iBSs,
the performance improvement from interference rank increase is smaller.
This is because multiple iBSs  spread the interference in space themselves.
Hence, further spreading interference into spatial subchannels does not 
derive as much
benefit as in case of single dominant interferer that uses single-layer
transmission.

\subsection{Discussion}

We have shown that higher rank transmission at an iBS may serve the UE of
interest better than a single rank transmission. Will it always lead to a better
performance?
An answer to that remains to be seen. We have provided one half of the story, that
is how the interference rank affects the UE of interest. We have shown that
there is potential to decrease outage of a weak link under strong interference,
especially if the weak link uses beamforming as its multi-antenna
technique. The other part of the story should consider the effect of the rank
choice on the own transmissions of the iBSs and evaluate the issue from a
system level perspective. We leave these thoughts for future consideration.

Another important issue here is how should our UE or sBS convey the request to
use higher transmission to one or more iBSs. Majority of LTE BSs are equipped
with the X2 interface to exchange control messages with other BSs and thus
could take advantage of it. However, not all BSs have this option. For
example, femto BSs are connected to the network via ADSL or similar last mile
connection that is not compatible with X2. In that case, a dedicated
over-the-air interface may be needed. In our opinion, trends in cellular
communications seem to be generally moving towards cooperative transmissions,
therefore considering an effect of transmission rank on a neighboring reception
should not be a major issue in the near future.

\section{Conclusions}
\label{sec:outro}

We have derived SINR distribution and probability of outage of beamforming
and OSTBC under arbitrary number of interferers with arbitrary transmission
power and several options of multi-antenna techniques. We have subsequently used
these to analyze impact of interference rank on a weak link that uses
beamforming or OSTBC as its transmission technique. Our results suggest that
the interference statistics of higher rank transmissions positively impact
performance by decreasing the probability of outage, leading to gain in the
supported SINR threshold.

\end{document}